\documentclass[12pt,a4paper]{article}
\usepackage{amsmath,amssymb}
\usepackage{graphicx,epsfig}
\usepackage{color}

\newcommand{\abs}[1]{\left\vert#1\right\vert}
\newcommand{\gsim}{\, \mathop{}_{\textstyle \sim}^{\textstyle >} \,}
\newcommand{\lsim}{\, \mathop{}_{\textstyle \sim}^{\textstyle <} \,}

\def\vev#1{\left\langle#1\right\rangle}

\begin{document}

\begin{titlepage}
\begin{flushright}
CERN-PH-TH/2005-162\\
UCB-PTH-05/25\\
LBNL-58803
\end{flushright}

\vskip 1.0cm

\begin{center}
  {\Large \bf Mirror World at the Large Hadron Collider}
 
   \vskip 1.0cm
 
  {\large Riccardo Barbieri$^{a,b}$, Thomas Gr\'egoire$^b$, Lawrence J.~Hall$^c$}\\[1cm]
{\it $^a$ Scuola Normale Superiore and INFN, Piazza dei Cavalieri 7, I-56126 Pisa, Italy} \\[5mm]
{\em ${}^b$Theoretical Physics Division, CERN, CH-1211 Gen\`eve 23, Switzerland}\\[5mm]
{\it $^c$ Department of Physics, University of California, Berkeley, and\\
  Theoretical Physics Group, LBNL,
  Berkeley, CA 94720, USA}\\[5mm]

     \vskip 1.0cm
 
   \abstract{A mirror world can modify in a striking way
the LHC signals of the Higgs sector. An exact or approximate $Z_2$ symmetry
between the mirror world and our world allows large mixing between the Higgs 
bosons of these worlds, leading to production rates and branching ratios for 
these states that are markedly different from the standard model and are 
characteristic of a mirror world.  The constraints on these Higgs boson 
masses from precision electroweak data differ from the standard model bound, 
so that the new physics that cancels the quadratic divergence induced by the top 
quark may appear at a larger scale, possibly beyond the reach of the LHC.  
However, the scale of new physics 
needed to cancel the quadratic divergence induced by the Higgs boson is not 
significantly changed. 
With small breakings of the $Z_2$ parity, the lightest mirror quarks (and 
possibly charged mirror leptons) could be the dark matter in the universe, 
forming galactic halos that are stable to cooling.
A possible signal from the relic radiation density of the mirror world is 
also discussed.}

\end{center}
\end{titlepage}
\def\baselinestretch{1.05}

\section{Introduction}

What signals of electroweak symmetry breaking are expected at the
Large Hadron Collider (LHC)? If a standard model (SM) Higgs boson of mass $m_h$ is
discovered, then the quadratic divergence in the Higgs mass squared
parameter from a top quark loop should be cutoff by new physics at a
scale
\begin{equation}
\Lambda_t \lesssim 400 \; \mbox{GeV} \left( {\frac{m_h}{115 \mbox{GeV}}}
\right){D_t^{1/2}},
\label{eq:LSM}
\end{equation}
where $D_t$ is the sensitivity  of the Higgs mass to $\Lambda_t$, or $1/D_t$ is the amount of fine tuning necessary to make $\Lambda_t$
large. Thus, if the Higgs is light, as suggested by electroweak
precision tests (EWPT), and the SM is completely natural, 
this new physics should be within reach of LHC. What is our expectation for 
this new physics?

The most successful and most ambitious answer to this 
question is the supersymmetric extension of the SM, as incorporated 
in the minimal supersymmetric standard model\cite{unif-mod, MSSM}. 
While smoothly passing the EWPT, this model succeeds in 
removing any significant cutoff dependence of the physical 
observables (apart from the cosmological constant) up to very high 
energies, even including the effects of the gravitational interactions. 
No other known model can claim a similar success. Yet, from the evolution 
of the experimental data, mostly from LEP2, another problem has emerged. 
The 
sensitivity of the Higgs mass parameter to the cutoff gets 
replaced by the sensitivity to the masses of the supersymmetric particles, 
so that, by applying the very same naturalness criterium to the simplest models, 
some of the superpartners, 
or the Higgs boson itself, should 
have already been discovered. Compared to the 
merits of the MSSM, this can be viewed as a relatively minor problem. 
Nevertheless it motivates exploring the consequences of a different, 
admittedly more modest point of view. 

Maybe the attempts at removing  any significant cutoff
dependence of the physical observables up 
to the highest possible energies is premature, given our  
relatively limited, direct experimental information above the Fermi scale. Perhaps one might be 
content with an extension of the SM which neatly keeps  the consistency 
with the EWPT or, in fact, with any other relevant experiment so far, 
but allows, at the same time, an increase of the naturalness cutoff 
by some amount, so that its very success does not look surprising (the 
``LEP paradox''\cite{Barbieri:2000gf}). This point of view has motivated a great part of the 
research on the electroweak symmetric breaking problem in recent years. 
As modest as it may be, the concrete realization of this program has proven 
not to be easy at all. 

A recent proposal in this direction\cite{Chacko:2005pe}
involves adding to the SM particles a mirror world and considering 
the Higgs as a pseudo-Goldstone boson of an approximate $SU(4)$ symmetry. 
In this paper we analyze the electroweak symmetry breaking in the mirror 
model in detail, paying special attention to
its naturalness properties and its experimental manifestations at 
the LHC. As we shall see, there is a possible  increase in $\Lambda_t$ 
relative to the SM bound of (\ref{eq:LSM}) that can be very significant 
for LHC. Such an increase can 
be achieved without fine-tuning and does not require
the Higgs to be a pseudo-Goldstone boson. 
In all parameter regions, even those with the highest possible
$\Lambda_t$, there are clear signatures at the LHC.

The entire SM is replicated in a mirror
world and a $Z_2$ symmetry interchanges our world with the
mirror world, ensuring identical particles and interactions. 
There are two renormalizable couplings that respect the mirror 
symmetry and connect the two worlds
\begin{equation}
\lambda_{12} H_1^\dagger H_1 H_2^\dagger H_2 ~~~~~~~~~~~~~ 
\epsilon B_1^{\mu \nu} B_2^{\mu \nu},
\label{eq:12int}
\end{equation}
where $H_{1,2}$ are the two Higgs doublets under the standard (1) and the mirror (2) gauge groups  respectively and $B_{1,2}$ are the hypercharge gauge bosons. The mixing induced by
$\lambda_{12}$ implies that the two mass-eigenstate Higgs bosons,
$h_\pm$, are non-degenerate 
and are non-trivial combinations of the components of $H_1$ and
$H_2$. This is the source of the LHC signals of the model \cite{Ignatiev:2000yy,Schabinger:2005ei}
.

Mirror symmetry was originally proposed as a way to restore
parity and has been investigated from many angles in recent years, as reviewed in \cite{Berezhiani:2003xm}, \cite{Foot:2003eq};  in particular, mirror baryons seem a natural
candidate for dark matter.  Nevertheless, mirror symmetry immediately introduces
several problems.
\begin{itemize}
\item The $B_1 B_2$  mixing term of 
  (\ref{eq:12int}) gives electric charges to mirror
  quarks and leptons, so that the coupling $\epsilon$ is constrained
  to be very small\cite{epsiloncharge}.
\item Equal baryon asymmetries in the two sectors implies equal
  amounts of baryonic matter and non-baryonic dark matter in the
  universe: $\Omega_{DM} = \Omega_B$.
\item The interactions of (\ref{eq:12int}) bring the two sectors into
  thermal equilibrium in the early universe, leading to 
an unacceptably large amount of radiation during the
eras of big bang nucleosynthesis (BBN) and cosmic microwave background
(CMB) generation.
\item Proto-galaxies of mirror matter will rapidly cool, as in the
  baryonic case, so that dark matter halos are unstable to collapse,
  unless a heating mechanism is introduced\cite{Foot:2004wz}.
\end{itemize}

Since the bounds on $\epsilon$ are very strong, in this paper we set
$\epsilon =0$ at some scale and we ignore the very small mixing between the photon and its mirror state reintroduced by radiative corrections, which is well below the current limits\footnote{The smallness of this mixing, vanishing at least up to and including 4 loops, can be traced back to the separate Charge Conjugation symmetries of the two worlds, violated only by exchanges among fermions of the intermediate W and Z bosons}.
To address the last three problems, which are cosmological, we introduce small
breakings of the mirror symmetry, giving sufficient mass to mirror quarks and charged leptons
to yield the observed amount of dark matter. This changes
the thermal history of the universe, greatly decreasing the radiation
energy in the mirror sector.  Furthermore, the mirror dark matter halo
is also stabilized.  

In Section 2 we consider the Higgs potential and its minimization. 
In Section 3 we discuss the naturalness properties of the model and the 
upper bounds on the cutoff scale. In Section 4 we outline the signals at 
LHC. Section 5 contains cosmological considerations concerning the 
interpretation of mirror baryons as dark matter and the mirror radiation 
energy density at BBN and CMB eras. A summary and conclusions are given in 
Section 6.

\section{The Higgs Potential }

In this section we study the Higgs potential, by first considering  the limit in which the Higgs 
potential is exactly invariant under the interchange of our world with
the mirror world\cite{Ignatiev:2000yy}.

The $Z_2$ invariant Higgs potential may be written in the form
\begin{equation}
V = - \mu^2 (H_1^\dagger H_1 + H_2^\dagger H_2) 
+ \lambda (H_1^\dagger H_1 + H_2^\dagger H_2)^2 
+ \delta [(H_1^\dagger H_1)^2 + (H_2^\dagger H_2)^2]. 
\label{eq:V}
\end{equation}
The $\lambda_{12}$ term
of (\ref{eq:12int}) is included in the term proportional to $\lambda$. As $\delta \rightarrow 0$,
the potential respects an $SU(4)$ symmetry. 
The vacuum of this theory has two phases, with $\delta = 0$ a critical point. 
We first briefly consider the case that one of the vacuum expectation
values (vevs) vanishes, and the remainder of the paper studies the
case that both vevs are non-zero.

\subsection{The Asymmetric Vacuum:  $\delta < 0$ }
For  $\lambda > |\delta|$ the potential is stable, and  only one Higgs boson 
acquires a vev. This non-zero vev must be in our sector, $\vev{H_1} = v \approx 175$ GeV with
\begin{equation}
v^2 = \frac{\mu^2}{2 (\lambda + \delta)},  ~~~~~~~~ \mbox{and} ~~~~~~~~~
m_h^2 = 4 v^2  (\lambda + \delta)
\label{eq:deltanegO}
\end{equation}
is the mass of our Higgs boson. The mass of the mirror sector Higgs
boson is
\begin{equation}
m_{H_2}^2 = 2 v^2  | \delta |.
\label{eq:deltanegM}
\end{equation}
The cubic Higgs interaction is 
\begin{equation}
\mathcal{L} = \frac{m_h^2 + 2 m_{H_2}^2}{\sqrt{2} v} \; h  H_2^2,
\label{eq:deltanegcubic}
\end{equation}
so that, if $\lambda > 3 | \delta |$,  $m_h > 2 {m_H}_2$ and our Higgs decays invisibly. 
This spontaneous breaking of mirror symmetry means that mirror QCD
will get strong at a lower scale than our QCD. However, assuming that
QCD with 6 light flavors spontaneously breaks its chiral symmetries,
the mirror quark condensates will break the mirror weak interactions. We do not
consider this phase further: the universe contains too much radiation
during BBN and the energy density in dark matter is less than that in baryons.

\subsection{The Symmetric Vacuum:  $\delta > 0$ }

If $\delta$ is positive the minimization
equations force both Higgs doublets to acquire the 
same vev, $\vev{H_i} = v_i$, with $v_1 = v_2 =v \approx 175$ GeV given by
\begin{equation}
v^2 = \frac{\mu^2}{2 (2 \lambda + \delta)}.
\label{eq:deltaposvev}
\end{equation}
The quantity $2 \lambda + \delta$ is positive for the potential to be stable.
The pattern of electroweak symmetry breaking is identical in the two
sectors, but because of the quartic interaction that mixes the two
Higgs doublets, the two mass-eigenstate neutral Higgs bosons have equal amplitudes
in the two sectors, $h_{\pm} =  (h_1 \pm h_2)/\sqrt{2}$, and have masses
\begin{equation}
m_+^2 = 2 \mu^2 = 4 v^2  (2 \lambda + \delta) ~~~~~~~ \mbox{and} ~~~~~~~ 
m_-^2 = 2 \mu^2 \frac{\delta}{2 \lambda + \delta} = 4 v^2  \delta.
\label{eq:deltaposHmasses}
\end{equation}

\subsection{$Z_2$ Breaking and Non-degenerate Vevs}

We now consider the potential
\begin{equation}
V = - \mu^2 (H_1^\dagger H_1 + H_2^\dagger H_2) + m^2 (H_1^\dagger H_1
- H_2^\dagger H_2) + \lambda (H_1^\dagger H_1 + H_2^\dagger H_2)^2 
+ \delta [(H_1^\dagger H_1)^2 + (H_2^\dagger H_2)^2] 
\label{eq:Vm}
\end{equation}
where $m^2$ is a $Z_2$ breaking mass parameter, and we ignore the $Z_2$
breaking quartic coupling. 
Requiring both vevs to be non-zero we find 
\begin{equation}
v_1^2 + v_2^2 =  \frac{\mu^2}{2 \lambda + \delta} ~~~~~~~\mbox{and}~~~~~~  
\frac{v_2^2}{v_1^2} = \frac{1 + y}{ 1 -y}
\label{eq:v2/v1}
\end{equation}
where
\begin{equation}
y = \frac{m^2}{\mu^2} \frac{2 \lambda + \delta}{ \delta},
\label{eq:y}
\end{equation}
so that $y$ must be between $-1$ and 1. With $\lambda$ and $
\delta$ of order unity, while  $\abs{m^2} \ll \abs{\mu^2}$ from small
$Z_2$ breaking,   $\abs{y} $ should be small and this vacuum is a
perturbation of the previous case. For example, the masses $m_\pm^2$ of 
(\ref{eq:deltaposHmasses}) are multiplied by $(1+y)$, and the $h_{1,2}$ 
components of $h_-$, important for the Higgs signals, deviate from $1/ \sqrt{2}$ 
by factors $(1 \pm \epsilon /2)$, with $\epsilon = ( 1+ \delta/\lambda)y$ and the 
positive sign applying to our sector. 

While the $Z_2$ breaking in the Higgs mass parameters, $m^2/\mu^2$, is
expected to be small, this could be counterbalanced by a small ratio
of quartics, $\delta/\lambda$. 
Hence we also
consider cases where $\abs{y}$ is not small.  When $y \rightarrow \pm 1$,  the vevs
become hierarchical, but this can be achieved only at the expense 
of a fine tuning, as discussed in the next section. 
Introducing $\tan{\theta} = v_1/v_2$, to leading order in
$\delta/\lambda$ the two mass eigenstate Higgs bosons are
\begin{equation}
h_+ = s \; h_1 + c \; h_2 ~~~~~~~\mbox{and}
 ~~~~~~~ h_- = c \; h_1 - s \; h_2
\label{eq:hpm}
\end{equation}
where $s= \sin{\theta}$ and $c= \cos{\theta} $, and  the masses are
\begin{equation}
m_+^2 = 2 \mu^2 ~~~~~~ \mbox{and}~~~~~~ m_-^2 = \mu^2  \frac{\delta}{\lambda} (1 - y^2).
\label{eq:mpm}
\end{equation}
Notice that if $\delta <0$, then $m_-^2 <0$ and the vacuum is
unstable. Thus, as in the $m^2=0$ case, $\delta = 0$ is a critical point: 
$v_1$ and $v_2$ both different from zero requires $\delta > 0$. 

\section{Naturalness}

With the vacua of the previous section we proceed to discuss the 
naturalness of given values of the observables,
considering first the case of exact $Z_2$ symmetry and later the 
$Z_2$ breaking effects from $m^2 \neq 0$.

\subsection{Naturalness for $m^2 = 0$.}

The quadratic divergence in the Higgs mass parameter induced at 1 loop
by virtual top quarks and their mirrors has the same form as in the SM:
\begin{equation}
\mu^2 = \mu_0^2 + a_t \Lambda_t^2,
\label{eq:mu2}
\end{equation}
where $\mu_0^2$ is the bare parameter,  $a_t = 3 \lambda_t^2/8 \pi^2$
and $\lambda_t = m_t/v_t \approx 1$.
One might therefore guess that the naturalness limit on the
scale of new physics, $\Lambda_t$, that cuts off this quadratic
divergence would be the same in the two theories. This is not the
case. 

In the SM, the physical Higgs boson mass is
\begin{equation}
m_h^2 = 2 \mu^2,
\label{eq:mh}
\end{equation}
so that the sensitivity of the Higgs mass to $\Lambda_t$ is given by
\begin{equation}
D_t(m_h) \equiv \frac{\partial \ln m_h^2}{\partial \ln \Lambda_t^2}
= a_t \frac{\Lambda_t^2}{\mu^2}.
\label{eq:Dt}
\end{equation}
Depending on how much
fine tuning is allowed, the new physics must be discovered at a scale
\begin{equation}
\Lambda_t 
= \frac{2 \pi}{\sqrt{3} \lambda_t} m_h D_t^{1/2}
\approx 400 \; \mbox{GeV} \left(
\frac{m_h}{115 \mbox{GeV}} \right) D_t^{1/2}.
\label{eq:Lambdat}
\end{equation}
Thus the LHC is expected to discover this new physics even if 
a 25\% fine tuning is allowed and $m_h$ is close to 285 GeV, the upper limit 
set at 95\% CL by the EWPT\cite{EWPT}. 

How does this change in the $Z_2$ symmetric mirror (M) theory? Since the
observables $v^2$ and $m_\pm^2$ are all proportional to $\mu^2$, and since
the quadratic divergence in $\mu^2$ is still described by
(\ref{eq:mu2}), they have the same sensitivity to the cutoff\footnote{
We ignore the mild logarithmic dependence of $\delta$ on $\Lambda_t$.}
\begin{equation}
D_t(v^2) = D_t(m_\pm^2)
= a_t \frac{(\Lambda_t^{M})^2}{\mu^2}
\label{eq:Dt2}
\end{equation}
when expressed in terms of $\mu^2$. However, whereas in the SM
 $\mu^2 = m_h^2/2$, in the mirror model it is given by $\mu^2 =
m_+^2/2$, so that
\begin{equation}
\Lambda_t^M =  \frac{2 \pi}{\sqrt{3} \lambda_t} m_+ D_t^{1/2}.
\label{eq:LtM} 
\end{equation}
The question of whether  $\Lambda_t^M$ is larger than  $\Lambda_t$ is
simply a question of whether $m_+$ is larger than $m_h$. How large can 
$m_+$  be taken?
Apparently $m_+$ can be made arbitrarily
large by increasing $\mu^2$, while lowering $\delta$ to ensure the
presence of a light Higgs boson, $h_-$. However, this violates bounds from
EWPT. In the SM, the relevant radiative corrections for EWPT involve $(\ln m_h)$ 
and lead to a
bound $m_h < m_{EW} \approx 285$ GeV. In the mirror model each mass
eigenstate Higgs contributes to the relevant radiative corrections,
but has a coupling to the SM sector which is $\sqrt{2}$
smaller than in the SM case. Hence the relevant radiative corrections
are obtained by the replacement 
\begin{equation}
\log{m_h} \rightarrow \frac{1}{2} \log{m_-} +  \frac{1}{2} \log{m_+},
\label{eq:mh2}
\end{equation}
so that the bound on the mirror model is $m_+ m_- < m_{EW}^2$. If
$m_+$ is made too large, then $m_-$ will violate the direct search
limit for the Higgs boson mass, which is about the same in the mirror model
as in the SM. Hence in the mirror model
\begin{equation}
\Lambda_t^M = \frac{2 \pi}{\sqrt{3} \lambda_t} m_-
\left( \frac{m_{EW}}{m_-} \right)^2   D_t^{1/2}
\label{eq:LTM}
\end{equation}
where we set $m_+$ to the largest value consistent with EWPT. 

Suppose that a Higgs boson is discovered with mass $m_{exp}$, identified with $m_h$ in the SM or with $m_-$ in the mirror model. Then, allowing
equal amounts of fine tuning in the two theories
and assuming that $m_+$ saturates the EWPT bound, gives
\begin{equation}
\frac{\Lambda_t^M}{\Lambda_t^{SM}} = 
\left( \frac{m_{EW}}{m_{exp}} \right)^2.
\label{eq:LtLt}
\end{equation}
Thus we see that a significant increase in the scale of the new
physics, $\Lambda_t^M$, is possible, although it depends sensitively on
$m_{exp}$. 
We stress that this increase arises exclusively
because the EWPT experimental bound on $m_+$ can be milder than 
that on $m_h$. As such, it  has little to do with interpreting $h_-$  
as an approximate Goldstone boson.
The increase in $\Lambda_t^M$ can  nevertheless be substantial and especially significant for the LHC: for $D_t$ in the range of 0.5 to 4,
and $m_{exp} = 120$ GeV, values of $\Lambda_t^{SM}$ range from 300 GeV 
to 800 GeV and are within reach of LHC. In the mirror theory, 
with the same values of $D_t$ and $m_{exp}$,
$\Lambda_t^M$ increases to 1.8 TeV -- 5 TeV, which could be outside the reach
of LHC, depending on the actual manifestation of the new physics at $\Lambda_t^M$.

The parameter $\mu^2$ also receives quadratically divergent radiative corrections from 
1 loop Higgs self-energy diagrams, leading to
\begin{equation}
\mu^2 = \mu_0^2 - a_H \Lambda_H^2,
\label{eq:mu2H}
\end{equation}
where $\Lambda_H$ is the scale at which new physics cuts off this divergence.
The sensitivity of $m_h$ to this scale is described by
\begin{equation}
D_H(m_h) \equiv \left| \frac{\partial \ln m_h^2}{\partial \ln
  \Lambda_H^2} \right|
= a_H \frac{\Lambda_H^2}{\mu^2}.
\label{eq:DH}
\end{equation}
In the SM $a_H = 3 \lambda/ 8 \pi^2$, while in the mirror model $a_H =
(5 \lambda + 3 \delta)/ 8 \pi^2$, leading to the scales
\begin{equation}
\Lambda_H^{SM} = \frac{4 \pi}{\sqrt{3}} v D_H^{1/2} \approx 1.3 \;
\mbox{TeV} \; D_H^{1/2}
\label{eq:LH}
\end{equation}
for the SM and
\begin{equation}
\Lambda_H^{M} = \frac{4 \pi}{\sqrt{3}} N v D_H^{1/2} \approx 1.4 \;
\mbox{TeV} \; D_H^{1/2}
\label{eq:LHM}
\end{equation}
for the mirror model. They differ only by the factor 
\begin{equation}
N = \frac{\Lambda_H^{M}}{\Lambda_H^{SM}} = \sqrt{\frac{6}{5} \left( 
\frac{\lambda + 0.5 \delta}{\lambda +0.6 \delta} \right)} 
\label{eq:N}
\end{equation}
which is close to unity. (The number quoted in (\ref{eq:LHM}) corresponds to the
case $\delta \ll \lambda$). Thus there is little change in the
naturalness prediction for the scale of new physics that cuts off the
quadratic divergence from the Higgs self energy.

\subsection{Naturalness for $m^2\neq 0$.}

Does the $Z_2$ breaking from $m^2 \neq 0$  
lead to significant changes in the scales $\Lambda_{t,H}^M$ from the
results of (\ref{eq:LTM},\ref{eq:LHM})?

Consider first the sensitivity of $m_+^2$ to variations in
$\Lambda_{t,H}$. Since  $m_+^2 = 2 \mu^2$, and the quadratic
divergence appears in $\mu^2$ as in (\ref{eq:mu2}) and  (\ref{eq:mu2H}),
$D_{t,H}^M$ have the same form as in  (\ref{eq:Dt2}) and  (\ref{eq:DH}),
so that, for small $\delta/\lambda$ and arbitrary $y$,
\begin{equation}
\Lambda_t^M =  \frac{2 \pi}{\sqrt{3} \lambda_t} m_+ D_t^{1/2}
~~~~~~~ \mbox{and} ~~~~~~
\Lambda_H^M = \frac{4 \pi v }{\sqrt{5}} \sqrt{1 + \frac{ v_2^2}{v_1^2}}
D_H^{1/2}
\label{eq:LtMLHM}
\end{equation}
For the divergence arising from the top quark, 
the crucial point is that the constraint from EWPT on how large
$m_+^2$  can be taken, is significantly changed.
Given the radiative corrections  in the SM, the corrections in the 
present theory are obtained by the replacement
\begin{equation}
\log{m_h} \rightarrow c^2 \log{m_-} + s^2 \log{m_+},
\label{eq:VPGB}
\end{equation}
implying that the current limit on the SM Higgs mass, 
$m_h < m_{EW} \approx 285$ GeV, gets replaced by
\begin{equation}
m_-^{c^2} m_+^{s^2} < m_{EW}
~~~~~ \mbox{or} ~~~~~
m_+ < m_- \left( \frac{m_{EW}}{m_-} 
\right)^{1 + \frac{v_2^2}{v_1^2}}.
\label{eq:EWPT2}
\end{equation}
As $v_2$ is increased above $v_1$, this bound  becomes in fact rapidly 
weaker than the perturbativity constraint from $\lambda < 4 \pi^2$:
\begin{equation}
m_+ < 4 \pi v \sqrt{1 + \frac{v_2^2}{v_1^2}}.
\label{eq:pertcon}
\end{equation}
 Taking, e.g., 
$v_2/v_1 =2$ and $m_- = 115 $ GeV, the bound from the EWPT, $m_+ < 11$ TeV, 
 exceeds  by far the perturbativity bound,  
$m_+ <  5$ TeV.
Note that a ratio $r = v_2/v_1$ significantly greater than 
unity, does require a fine tuning: $r = \sqrt{2} D_r^{1/2}$,
where $1 / D_r$ parameterizes the amount of fine tuning needed to make
$r$ large, or
\begin{equation}
v_2 \approx 250 \; \mbox{GeV} \; D_r^{1/2}.
\label{eq:v2finetune}
\end{equation}
Nevertheless, even moderate values for $v_2/v_1$ lead to substantial
increase in $\Lambda_t^M$.
For example, taking $D_r = 2$, hardly a fine tune, and $\lambda = \frac{\pi^2}{2}$, 
well inside the perturbative domain, gives $m_+ = 1.7$ TeV and
$\Lambda_t^M = 6.2$ TeV $D_t^{1/2}$ for any $m_-$ below 180 GeV.

The observables $m_-^2$ and
\begin{equation}
v_{(2,1)}^2  = \frac{\mu^2}{4 \lambda}\pm  \frac{m^2}{2 \delta}
=  \frac{\mu^2}{4 \lambda} (1 \pm y)
\label{eq:v21}
\end{equation}
have sensitivities $D_{t}$ to quadratic divergences
which, for $y$ close to $1$, are a factor of $1/(1-y)$ larger than those
for the observable $m_+^2$. However, this factor is too mild to affect
the above arguments.
Hence, for $m_\pm$ that satisfy the EWPT bound,    
there is a completely natural region of parameters
that pushes the physics responsible for cutting off
the quadratic divergence from the top quark beyond the reach of
LHC. 

The natural value for the scale that cuts off the
quadratic divergence arising from Higgs boson loops is unchanged from
the $Z_2$ symmetric case of the last sub-section  
\begin{equation}
\Lambda_H^M = \sqrt{\frac{2}{5}} 4 \pi v (D_H D_r, D_H)^{1/2}
\approx 1.4 \; \mbox{TeV}  (D_H D_r, D_H)^{1/2}.
\label{eq:LHM2}
\end{equation}
The sensitivity factor $D_H D_r$ applies for the operators $m_+^2,
v_2^2$, while the factor $D_H$ applies for the operators  $m_-^2,
v_1^2$. 
Thus the mirror sector makes no substantial change from the SM for the
scale expected for this physics.

In the above arguments we have assumed that the top Yukawa couplings
and the Higgs quartic couplings are $Z_2$ invariant, so that the
quadratic divergences only affect $\mu^2$. Relaxing this assumption
leads to a quadratic divergence also for $m^2$, which affects
$v_{2,1}$, as seen from (\ref{eq:v21}). Suppose that $\mu^2 = \mu_0^2 +
a_\mu \Lambda^2$ and  $m^2 = m_0^2 + a_m \Lambda^2$, for either the $\Lambda_t $ 
or the $\Lambda_H$ case. Then the relative
sensitivity coming from the divergence in $m^2$ to that in $\mu^2$ for
the observable $v_1^2$ is
\begin{equation}
 \frac{D^M(v_1^2)|_m}{D^M(v_1^2)|_\mu}
= \frac{a_m}{a_\mu} \frac{m_+^2}{m_-^2} (1 - y^2),
\label{eq:Dv1}
\end{equation}
showing that the possibility of naturally increasing $\Lambda_t$ 
relative to the SM rests on a small $a_m / a_\mu$, ie on a small 
violation of the $Z_2$ symmetry in the large dimensionless couplings of
the theory.  To increase $\Lambda_t$ it is not sufficient to mix the Higgs
doublet with any SM singlet scalar --- the mirror world plays a crucial role.

\subsection{The Strong Coupling Limit}

In the case of strong Higgs self coupling, $\lambda \approx 4 \pi^2$,
our 1 loop Higgs analysis cannot be justified.
One can nevertheless estimate $m_+  \simeq 4 \pi v  \sqrt{1 + \frac{v_2^2}{v_1^2}}$.
 For an estimate of $\Lambda^M_H$ we take the scale at which the $W_L
 W_L$ scattering amplitude saturates the unitarity limit. Indeed, due 
to an incomplete cancellation from the light Higgs exchange diagram, 
for a c.o.m. energy $\sqrt{s}$ below $m_+$, the $W_L W_L$ scattering 
amplitude grows as 
\begin{equation}
T( W_L W_L \rightarrow W_L W_L) = - \frac{G_F}{\sqrt{2}} s (1 + \cos{ \theta_S}) \frac{v_1^2}{v_2^2 + v_1^2},
\label{eq:VPGB2}
\end{equation} 
where $\theta_S$ is the scattering angle,
and similarly for $Z_L Z_L$. Therefore, 
the unitarity bound  is saturated at
\begin{equation}
s_c = \left( \frac{v_2^2 + v_1^2}{v_1^2} \right) s_c^{SM}
 \label{eq:bound}
\end{equation}
where $s_c^{SM} $ is the analogous SM bound, $s_c^{SM} = (1.2 \; \text{
TeV})^2$ from the most constraining isospin zero channel\cite{Lee:1977eg}. 
Fine tuning $D_r$ large to increase $v_2$ gives  a bound
$ \Lambda^M_H \simeq \sqrt{s_c}  \simeq 1.7 \; \mbox{TeV} D_r^{1/2} $,
which is certainly consistent with the perturbative bound of (\ref{eq:LHM2}). 
Since the top quark is perturbative, $\Lambda_t^M$ is about a factor of 
$\pi$ larger.

In this strong coupling limit $\lambda \gg g, \lambda_t, \delta \approx 1 $, 
and hence the Higgs potential possesses an approximate $SU(4)$ symmetry.
This case was the focus of  Chacko et al\cite{Chacko:2005pe}, and the Higgs boson
may be viewed as a pseudo-Goldstone boson. In this limit, we note that, since
$\lambda$ becomes non-perturbative, two-loop diagrams
give radiative contributions to $\delta$ that are of order $ g^2$.
This suggests that $m_-$ is likely to be closer to its upper limit of $m_{EW}$
rather than its lower limit of 115 GeV, since the latter requires a modest
fine-tune.

\section{LHC Signals}

The quartic coupling $\lambda$ of the Higgs potential connects the
SM and mirror sectors, so that the mass eigenstate Higgs bosons are
the linear combinations $h_\pm$ shown in (\ref{eq:hpm}), with masses
$m_\pm$ of (\ref{eq:mpm}).  The couplings of these mass-eigenstate Higgs bosons to SM fermions, $\psi_1$, and to mirror
fermions, $\psi_2$, are
\begin{equation}
\lambda_1 \; \overline{\psi_{1_{L}}} \psi_{1_{R}} (c \; h_- + s \; h_+)
+\lambda_2 \; \overline{\psi_{2_{L}}} \psi_{2_{R}} (c \; h_+ - s \; h_-)  + h.c.
\label{eq:hcoup}
\end{equation}
The Yukawa couplings $\lambda_1$ are numerically identical to those of
the SM and, in the $Z_2$ limit, $\lambda_2 = \lambda_1$. In
the next sections we show that $Z_2$ breaking must be significant for
the light quark and lepton Yukawa couplings, so that for these species
$\lambda_2 > \lambda_1$. As before, we ignore any small $Z_2$ breaking in
the top quark Yukawa coupling and in the gauge couplings.

\subsection{Small $y$}

Since $Z_2$ breaking effects are small, it is natural to expect that
$y$ is small and that the electroweak vevs of the two sectors are
nearly equal. In this limit the Higgs mixing angle is $45^o$, resulting
in a factor of $1/\sqrt{2}$ in the coupling of $h_\pm$ to each sector\cite{Foot:1991py},\cite{Ignatiev:2000yy}.
The three free parameters of the Higgs potential are determined in
terms of $m_\pm$ and the electroweak vev $v$, giving a predictive
phenomenology. 

Production of either $h_+$ or $h_-$ at the LHC is suppressed by a
factor 2 relative to the production rate of the SM Higgs, $h$, 
\begin{equation}
\sigma ( i \rightarrow  h_{\pm} +X) = \frac{1}{2} \sigma ( i \rightarrow  h +X)
\label{eq:prod}
\end{equation}
with $m_h = m_{\pm}$.
The same suppression factor applies at LEP, but the direct
search bound, $m_\pm \gsim 110$ GeV, is little changed from the
SM, even if the decay branching ratio to invisible
particles is substantial.  The EWPT bound, $m_+ m_- < m_{EW}^2$,
implies that both $h_+$ and $h_-$ will be accessible at LHC.
As argued in the previous section, an increase of $\Lambda_t$ above the 
SM value occurs if $m_\pm$ are hierarchical and the EWPT bound is saturated.

Concerning the branching ratios, let us first consider the lightest state, $h_-$.
Every width into a visible decay mode is reduced by a factor 1/2 relative to the same width of the SM model Higgs with $m_h = m_-$. At the same time all these widths are identical to the widths into the corresponding invisible mirror mode, except for the effects  of $Z_2$ breaking in the 
 light quark and lepton Yukawa couplings.
Therefore, all branching ratios into
visible  modes of $h_-$ are reduced compared to the
corresponding SM case by a factor $1 + f$, where $f$ is the ratio
of decays to mirror states compared to SM states,
\begin{equation}
BR (h_-  \rightarrow  X) = \frac{1}{1+f} BR (h  \rightarrow  X),
\label{eq:BR}
\end{equation} 
and the event rates at LHC for $\bar{b} b, \bar{\tau} \tau$ or $\gamma \gamma$ from $h_-$-decays
are a factor $2(1+f)$ smaller than the rates for a SM Higgs
of the same mass.
The factor $1 + f$ is universal and depends only on $m_-$. 
If the $W^+ W^-$ channel is open,  $f$ is close to unity. 
On the contrary, for $m_-$ below $2 M_W$, 
a relatively large violation of $Z_2$ symmetry in the quark and
lepton Yukawa couplings other than the top one could lead to values of $f$ far from unity. From BBN
considerations, with $m_- < 2 M_W$, we expect $f$ greater than unity, so that invisible
decays to mirror particles become the dominant decay mode.

If $m_+ < 2m_-$, all the results of the previous paragraph
also apply for the LHC signals of $h_+$. 
However,  if $m_+ > 2 m_-$ then the decay mode $h_- h_-$ must be
included\footnote{ Note that $\lambda < 0$ reverses the order of the masses, $m_+ < m_-$. This would give rise to a different phenomenology, since the coupling $h_- h_+ h_+$ vanishes in the exact $Z_2$-symmetric limit. 
We have not emphasized this case because $\Lambda_t$ cannot be increased 
above the SM value.}. The cubic 
Higgs interaction 
\begin{equation}
\mathcal{L} = \frac{m_+^2 + 2 m_-^2}{4v} \; h_+  h_-^2
\label{eq:deltaposcubic}
\end{equation}
leads to the decay rate
\begin{equation}
\Gamma (  h_+  \rightarrow  h_- h_-) = \frac{m_+^3}{128 \pi v^2} (1 +
2 x)^2 
\cdot (1 - 4 x)^{1/2},
\label{eq:decay}
\end{equation}
where $x = m_-^2/m_+^2$.
The $h_+$ branching ratios to $W^+ W^-, \bar{t}t$ and $h_- h_-$ are
shown in Figure 1 as a function of $m_+$, with $m_-$ chosen to be 110
GeV. For $2 M_W < m_+ < 2 m_-$, half of all decays are to  $W^+ W^-,
ZZ$. These channels continue to dominate the visible decays at higher
$m_+$, although the $h_-h_-$ mode is not much smaller. Once $m_\pm$
are both measured, these branching ratios are precise predictions of
the theory and, if confirmed, would give striking evidence in favor of
an entire mirror sector.

Finally, a property of $h_-$ that neatly distinguishes it from the SM Higgs is the vanishing of its triple self-coupling, as required by $h_- \rightarrow - h_-$ under the $Z_2$ symmetry\footnote{ See \cite{Dahlhoff:2005sz} and references therein for the prospects to measure the trilinear self-coupling and other Higgs couplings at the LHC.}.

\begin{figure}[t!]
\begin{center}
\includegraphics[width=10cm]{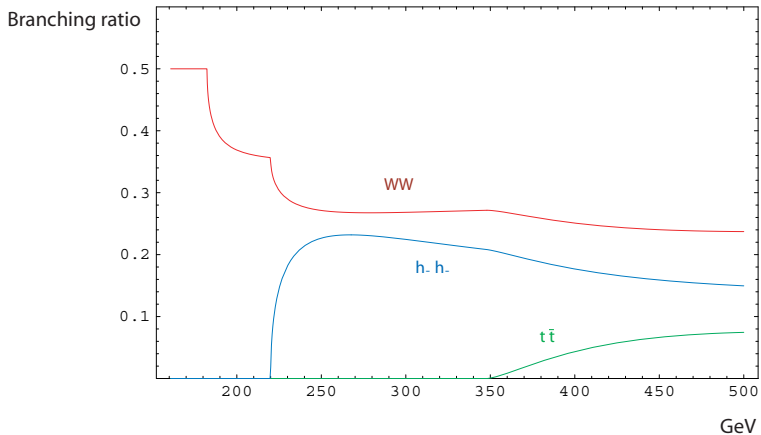}
\end{center}
\caption{\label{Br} Branching ratio of the heavy Higgs, as function of its mass in GeV, to $WW$ (in red), $t \bar{t}$ (in green)  and $h_- h_-$ (in blue) with $m_- = 110 \text GeV$}
\end{figure}

\subsection{Increasing $y$}
By increasing $y$, from (\ref{eq:v2/v1}) the ratio $v_2/v_1$ deviates
from unity. The main consequence of this is the change in the
composition of the $h_\pm$ states, eq. (\ref{eq:hpm}), and the 
consequent weakening of the EWPT bound on $m_+$, eq. (\ref{eq:EWPT2}). 

The coupling
of the light Higgs to known particles are reduced by a factor $c$ compared to
those of the SM. Precision studies of the Higgs may be required to uncover
this. The heavier Higgs, $h_+$, is coupled to SM particles through a factor $s$ and its coupling to $h_- h_-$ is obtained from  (41)  by sending $v$ into ($1/\sqrt{2}) (v_1^2 + v_2^2)^{1/2}$. An increase of $m_+$ can make the detection of  $h_+$ 
difficult at LHC. However, as recalled in Sect. 3.3, 
at a c.o.m. energy $\sqrt{s}$ below $m_+$ the $W_L W_L$ scattering 
amplitude grows as in eq. (\ref{eq:VPGB2}).

As a consequence of all this, the evidence of the signals described in 
the previous sub-section gets diluted. On the other hand, as noticed 
in Sect. 3.3 , a ratio $v_2/v_1$ significantly greater than unity 
requires a corresponding increase in the amount of fine tuning.

\section{Cosmology}

How is it possible for the mirror radiation not to violate BBN 
and CMB bounds, while having the mirror baryons giving sufficient dark matter?
Most attempts have assumed that the temperature of the two sectors differ 
after inflation, and that the cosmological baryon asymmetry in the mirror 
sector is larger than in our sector 
\cite{Berezhiani:1995am,Foot04,Berezhiani}.
Here we pursue an alternative where the two temperatures equilibrate
and the two baryon asymmetries are identical.

\subsection{Mirror Baryons as Dark Matter}

The lightest mirror quarks and leptons are an intriguing candidate for
dark matter (DM). For $\Omega_{DM} > \Omega_B$ the $Z_2$ parity must be
broken, by raising the mirror baryon mass, $m_{B_2} > m_{B_1}$,  and/or 
the mirror cosmological baryon
asymmetry, $\eta_{B_2} > \eta_{B_1}$. In the hope of obtaining
predictions, we assume  $\eta_{B_2} = \eta_{B_1}$.

An asymmetry in the light quark and lepton masses of the two sectors
may arise if there are terms in the  fermion mass matrices
proportional to powers of flavon fields $\phi_{1,2}$
with asymmetric vevs. This can happen very easily: the potential
for the flavons will have the form of (\ref{eq:V}), with $H_{1,2}$
replaced by $\phi_{1,2}$, and if $\delta<0$ an asymmetric vacuum with
$\phi_1 =0$ and $\phi_2 \neq 0$ can result, giving masses only to the
mirror fermions and leading to a larger energy density in CDM than in
baryons. Of course, other flavons will have vevs in the SM sector, 
giving mass to SM quarks and leptons as well. 

The precise nature of the mirror DM particle depends on the
spectrum of the light charged mirror fermions.
For example, suppose
that the lightest states are $u_2, d_2$ and $e_2$, and that the heaviest of
these, which must have a mass larger than the mirror QCD scale, 
is able to beta decay into the other two. Depending on which is
heaviest, the DM particle is  $N_2^0(u_2d_2d_2), \Delta_2^-(d_2d_2d_2)e_2^+$
or $\Delta_2^{++}(u_2u_2u_2)e_2^- e_2^-$.
Here, $\pm$ refers to mirror electric charge, which is necessarily 
unbroken, and therefore has no cosmological asymmetry. 
Furthermore the mixing of hypercharge gauge bosons is 
sufficiently small to be irrelevant in both the laboratory and in cosmology.
The mass of  $N_2^0, (\Delta_2^-e_2^+)$ or $(\Delta_2^{++}e_2^- e_2^-)$ must be about 5 GeV, but this leaves a wide range for the underlying
spectrum. We denote the lightest mirror quark or charged lepton mass by 
$m_<$ and the next-to-lightest by $m_>$. In the degenerate case, $m_>
\approx m_< \approx$ 1 -- 2 GeV, but in the hierarchical case $m_<$
could be very light and $m_>$ could be as large as 5 GeV.  An
important advantage of breaking mirror symmetry in the light fermion
spectrum is that this prevents cooling and collapse of DM
halos. Stability of the halos from mirror bremstrahlung or ionization  
cooling is automatic, as long as  $m_{e_2} \gsim 10^{-2} m_>$.

The virtual exchange of Higgs scalars, $h_\pm$, will allow mirror DM 
particles to interact in detectors searching directly for DM particles
impinging on Earth. However, the interaction cross section on protons,
$\sigma \approx 10^{-46} \div10^{-45} \mbox{cm}^2$ for $m_- = 110$ GeV, is far too
small for observation at present  detectors. Furthermore,
since the DM particle mass is 5 GeV, the recoil energy is low. The
accessible signals for this mirror DM lie in the relic radiation
energy density.

\subsection{The Mirror Radiation Energy Density at BBN and CMB Eras}

The quartic interaction $H_1^\dagger H_1 H_2^\dagger H_2$ plays a
crucial role in the thermal history of the universe.
At high temperatures, of order the weak scale, it ensures that the
mirror and SM sectors are thermally coupled at a single temperature. 
At some lower temperature, $T_M$, the mirror sector loses thermal
contact with the SM particles. As the universe further expands and
a SM (mirror)
quark or lepton mass threshold crossed, annihilations will reheat 
only the SM (mirror) sector. The non-degeneracies between the light
fermions of the two sectors, required for sufficient CDM as discussed
above, will then lead to a temperature asymmetry.  Only those species with
masses less than $T_M$ can lead to a temperature asymmetry, and since 
the mirror fermions are heavier than their SM partners, $T_1 > T_2$.
The theory predicts a correlation: the sector that contributes
more non-relativisitic matter today will have less
radiation -- as observed. Is the smaller amount of relic radiation 
in the mirror sector compatible with observations of BBN and the CMB?

The temperature $T_M$ depends on the spectrum of the mirror sector and
the mass of $h_-$. Given the constraints on the spectrum from the
requirements of mirror dark matter abundance, and the experimental
constraint 110 GeV $< m_-<$ 230 GeV, we find
\begin{equation}
2 \; \mbox{GeV} \lsim T_M \lsim 6 \; \mbox{GeV}.
\label{eq:TM}
\end{equation} 
To minimize the mirror radiation component, we assume that the only
mirror particles with mass below $T_M$ are the mirror
photon, gluons and neutrinos. In this case we find that the mirror
sector contributes an effective number of neutrino species to the
radiation of $\Delta N_\nu \simeq 1.4$.  While this result is disfavored
by the inferred primordial abundances of light elements, it will be
definitively tested by the PLANCK satellite.  

This prediction follows from simple thermodynamic arguments, but it
does have important caveats.  It would be changed if the QCD phase
transitions in the SM and mirror sectors generate significant entropy.
The phase transitions are very different --- in the SM there are 2 very
light flavors whereas in the mirror case there are one or zero --- so that the
amount of entropy generated could be very different \cite{Chacko:2005pe}. 
A second caveat
is that we have assumed that none of the entropy of the mirror gluons 
leaks into the SM sector.  This may require a more accurate analysis
in the case that $T_M$ is near its lower bound of 2 GeV, as the mirror
QCD scale may be as much as a factor 2 or 3 larger than the SM QCD
parameter. 

\section{Summary and Conclusions}

The SM Higgs sector involves a single unknown parameter, the Higgs
mass, which is constrained by EWPT: $m_h < 285$ GeV at 95 \% CL. The Higgs sector
of a mirror world with an approximate $Z_2$ symmetry is expected to
have just two unknown parameters, the masses $m_\pm$ of the two Higgs
bosons $h_\pm$. The production and decay of these states is completely 
determined by these parameters. In particular, production rates are suppressed by a
characteristic factor of 2, and the leading visible
decay branching ratios for $h_+$ are shown in Figure 1. The LHC could
convincingly demonstrate the existence of the mirror sector.  If $h_-$
is below the $WW$ threshold, its visible decay modes may be sensitive
to small $Z_2$ breaking effects.  
The scale $\Lambda_t$ at which new physics should damp the quadratic
divergence originating from the top quark is proportional to $m_+$ in
this theory, rather than $m_h$, as in the SM.
Since EWPT place an upper bound on the product $m_+ m_-$, if $h_-$ is light 
$m_+$ may be heavy, allowing $\Lambda_t$ to be up to a factor 6 larger
than in the SM.

While the $Z_2$ breaking in the Higgs mass parameters, $m^2/\mu^2$, is
expected to be small, this could be counterbalanced by a small ratio
of quartics, $\delta/\lambda$. In this special situation the Higgs vevs
of the two sectors are no longer equal, so that  Higgs phenomenology now
depends on the additional parameter $v_2/v_1$. A large value for
$v_2/v_1$ requires a fine tune and hence is disfavored;  but an
intriguing aspect of this region of parameter space is that, even
without fine tuning, moderate values of $v_2/v_1$ largely remove the
EWPT bound on $m_+$, so that $\Lambda_t$ can be increased well beyond
the reach of LHC. This can occur even if $m_-$ is significantly above
the limit from direct searches.
If this new physics at $\Lambda_t$ is not seen at
LHC, then it will be important to check for the deviations of $h_-$
phenomenology from that of the SM Higgs and to search for the heavier state
$h_+$, or, in the limit that the quartic $\lambda$ is strongly
coupled, to search for strong $WW$ scattering which should saturate the
unitarity limit at center of mass energies of about 2 TeV.

Small $Z_2$ breaking effects may be large enough in the
light quark sector to give stable mirror baryons a mass 5 times larger
than the proton mass, generating sufficient mirror baryon dark matter 
even if both sectors have the same baryon asymmetry. Such $Z_2$
breaking allows dark matter halos to be stable against cooling, and
also reduces the relic mirror radiation to $\Delta N_\nu \approx 1.4$,
in the absence of entropy generation from the two QCD phase transitions.

\section{Acknowledgments}

We are grateful to Z. Chacko and Riccardo Rattazzi for several useful discussions.
This work is supported by the EU under RTN contract
MRTN-CT-2004-503369. The work of R.Barbieri is also supported by MIUR.
The work of L.J.Hall was supported by the US Department of Energy under
Contract DE-AC03-76SF00098 and DE-FG03-91ER-40676, and by the
National Science Foundation under grant PHY-00-98840.

\end{document}